\documentclass[twoside,7pt]{amsart}
\usepackage{amsmath,amssymb,mathrsfs}
\def\d{\operatorname{d}}\def\<{\langle}\def\>{\rangle}
\def\Tr{\operatorname{Tr}}\def\:{\hbox{\bf :}}
\def\Cmplx{\mathbb C}\def\Intgrs{\mathbb Z}
\def\map#1{{\mathscr{#1}}}

\def\set#1{{\sf #1}}\def\alg#1{{\mathcal #1}}\def\aA{\alg{A}}
\def\aH{\alg{H}}
\def\grp#1{{\mathbf #1}}
\def\transp#1{{#1}^\tau}

\def\dual#1{{#1}^\tau}
\def\rnk{\operatorname{rank}}
\def\Rng{\set{Rng}}\def\Ker{\set{Ker}}
\def\Supp{\set{Supp}}
\def\Klm{{\mathfrak X}}
\def\dim{\operatorname{dim}}
\def\Span{\set{Span}}
\def\spc#1{{\mathscr{#1}}} 
\def\T#1{\alg{T_1(\spc{#1})}}
\def\Bnd#1{\alg{B(\spc{#1})}}\def\Bndd#1{\alg{B}(#1)}\def\St#1{\alg{T_1^+(\spc{#1})}}
\def\sH{\spc{H}}\def\sK{\spc{K}}
\def\sA{\spc{A}}\def\sB{\spc{B}}\def\sS{\spc{S}}\def\sL{\spc{L}}

\def\gG{\grp{G}}\def\gI{\grp{I}}
\def\qed{$\,\blacksquare$\par}
\def\ie{i. e. }\def\SU#1{\mathbb{SU}(#1)}\def\U#1{\mathbb{U}(#1)}

\def\dag{\dagger}
 \newcommand{\ket}[1]{ | \, #1  \rangle}
\newtheorem{lemma}{Lemma}

\newtheorem{corollary}{Corollary}
\newtheorem{theorem}{Theorem}
\def\Proof{\medskip\par\noindent{\bf Proof. }}
\def\example{\subsection{Example}}
\begin{document}
\title{Extremal covariant quantum operations and POVM's}
\author{Giacomo Mauro D'Ariano}
\email{dariano@unipv.it}
\address{{\em QUIT} Group, http://www.qubit.it, Istituto Nazionale di Fisica della Materia,
Unit\`a di Pavia, Dipartimento di Fisica ``A. Volta'', via Bassi 6,
I-27100 Pavia, Italy, and \\ Department of Electrical and Computer
Engineering, Northwestern University, Evanston, IL  60208}
\date{\today}
%\pacs{03.65.-w 03.65.Ta 03.67.-a 03.65.Fd 02.20}
\begin{abstract}
We consider the convex sets of QO's (quantum operations) and POVM's
(positive operator valued measures) which are covariant under a general
finite-dimensional unitary representation of a group. We derive 
necessary and sufficient conditions for extremality, and give general
bounds for ranks of the extremal POVM's and QO's. Results are
illustrated on the basis of simple examples.   
\end{abstract}
\maketitle
\section{Introduction}
The need for miniaturization and the new quantum information
technology\cite{Nielsen} has recently motivated a search for
new quantum devices with maximum control at the quantum
level. Among the many problems posed by the new technology there is
the need of engineering quantum devices which perform specific
measurements \cite{Holevo82,Helstrom,Davies,Busch} or 
particular state transformations---the so-called {\em quantum
operations} \cite{Kraus83a,davies-book,Stinespring}---which are optimized
with respect to some given criterion.  In most cases such optimal quantum  
measurements/operations are {\em covariant}\cite{Davies70} with respect to
a group of physical transformations. For the case of a quantum
measurement, "group-covariant" means that there is an action of the
group on the probability space which maps events into events, in such
a way that when the quantum system is transformed according to a
group transformation, the probability of the given event
becomes the probability of the transformed event. This situation is
very natural, and 
occurs in most practical applications. For example, the heterodyne
measurement\cite{YuenSha,Bilk-poms} is covariant under the group
of displacements of the complex field, which means that if we displace the state of
radiation by an additional complex averaged field, then the output
photo-current will be displaced by the same complex quantity. 
\par In quantum mechanics the probabilities for a given apparatus for all
possible states are described by positive operator valued measures
(POVM)\cite{Helstrom}, and we will say that the measurement is covariant when
its POVM is covariant under a unitary group representation\cite{note1,Holevo82}. 
For quantum operations (QO), on the other   
hand,  covariance means that the output of a group-transformed input state is 
simply the transformed output state---a situation
again quite common in practice. Typically covariance means that the apparatus is
required to work equally well on a full set of states which is
invariant under a group of transformations. For
instance, if one wants to engineer an eavesdropping apparatus for a
BB84 cryptographic scheme \cite{BB84,gisirev} that 
clones equally well all equatorial qubits, then the optimal cloning
operation must be covariant under the group $\gG=\Intgrs_4$ of $\pi/2$
rotations of the Bloch sphere around its polar axis, which is a
subgroup of the group of all axial rotations $\gG=\U{1}$\cite{opt_pcc}. Similarly, if
one wants to engineer a QO which works 
equally well on all pure states, then the operation must be covariant
under the full $\SU{d}$ group, where $d$ is the dimension of the
Hilbert space of the quantum system. 
\par It is easy to see that all POVM's covariant under some group
representation make a convex set, which describes the complete class of possible
covariant apparatuses. The same obviously holds for 
group-covariant QO's. Typically in most applications the
optimization resorts to minimize a concave function on the convex set
of covariant machines (in quantum estimation theory\cite{Helstrom}
actually such function is generally linear), whence the optimal
machine will correspond to an extremal element of the convex set. For
such purpose it is convenient to classify all extremal covariant
POVM's and QO's, and this is precisely the subject of
the present paper.
\par For finite dimensional Hilbert space, a characterization of all
non-covariant extremal QO's was given in
Ref. \cite{Choi}, whereas a characterization of all  extremal POVM's
can be found in Refs. \cite{Parthasaraty} and \cite{extrpovm} for
discrete finite probability space. On the other hand, no
classification of the extremal QO's or POVM's is
available yet under a covariance constraint, since, as we will see,
this constraint makes the classification 
problem much harder. Coincidentally, in many
applications the optimal QO/POVM is 
restricted to be rank-one from the special form of the optimization
function (this is the case, for example, of optimal phase estimation
for pure states\cite{Holevo82,Helstrom,pomph}, or of  phase covariant
optimal cloning of pure states\cite{opt_pcc}), and this has lead
to a widespread belief that optimality is synonym of
rank-one. However, as we will see in this paper, for sufficiently  
large dimension the extremal QO's/POVM's can easily 
have rank larger than one: this can actually happen for
optimization with mixed input states, such as in the case of
optimal phase estimation with phase-coherent mixed
states\cite{arrowoftimephase}.  
\par In this paper we provide a classification for finite dimensions of
all extremal POVM's and QO's that are covariant under
a general unitary group representation. We will generally consider continuous Lie
groups, since then all results will also apply to the case of discrete
groups as well, with just a little change of notation. We provide
necessary and sufficient conditions for extremality, along with simple
necessary conditions, which allow to "sieve" the extremal
QO's/POVM's. From these conditions general bounds for the rank
of the extremal QO's/POVM's easily follow as corollaries.
\par The paper is organized as follows. In Sect. \ref{POVM}
we briefly review the concept of POVM and that of covariant POVM based
on the Holevo's theorem\cite{Holevo82}. In Section \ref{qo} we recall
the necessary concepts about QO's, including their
operator form introduced in Ref. \cite{clon}, which allows to easily
classify the covariant QO's as non-negative operators in
the commutant of a suitable representation of 
the group. Section \ref{techlemma} is entirely devoted to some technical
lemmas which will be used in the classification of both POVM's and
QO's. Finally Sections \ref{extrcovP} and \ref{excovqo}
contains the classification theorem of extremal group covariant POVM's and
QO's, respectively, with some simple explicit  
examples, in particular with application to phase-covariant
estimation and phase-covariant optimal cloning.   
\section{Positive operator valued measures (POVM)}\label{POVM}
In the following we will denote by $\Bnd{K,H}$ the linear space of
bounded operators from the Hilbert space $\sK$ to the Hilbert space
$\sH$, and by $\Bnd{H}\doteq\Bnd{H,H}$ the algebra of bounded
operators on $\sH$. By $\T{H}$ we will denote the trace-class
operators on $\sH$, and by $\St{H}$ its positive elements.
\par A general measurement is described by a probability space $\Klm$
equipped with a sigma-algebra structure $\sigma(\Klm)$ of measurable
subsets $B\in\sigma(\Klm)$. The measurement returns a
random outcome $x\in\Klm$. In quantum mechanics the
probability that the outcome belongs to a subset $B\in\sigma(\Klm)$ 
depends on the state $\rho\in\St{\sH}$ of the system in a way which is 
distinctive of the measuring apparatus according to the Born rule
\begin{equation}
p(B)=\Tr[P(B)\rho],\label{Born}
\end{equation}
where $P$ is a function on $\sigma(\Klm)$ which is positive-operator
valued in $\Bndd{\sH}$, with the normalization condition
\begin{equation}
P(\Klm)=I_\sH.\label{normP}
\end{equation}
Positivity of $P$ is needed for positivity of probabilities for every
state $\rho$, whereas Eq. (\ref{normP})  guarantees
normalization of probabilities. In synthesis, $P$ is a 
positive operator valued measure (POVM) on the probability space
$\Klm$. In a sense the POVM $P$ represents our
knowledge of the measuring apparatus from which we can infer
information on the state $\rho$ from probabilities. The linearity of
the Born rule (\ref{Born}) in both arguments $\rho$ 
and $P$ is consistent with the intrinsically statistical nature of the
measurement, in which our partial knowledge of both the system and 
the apparatus reflects in convex structures for both states and
POVM's.  This means that not only states, but also POVM's can be
"mixed", namely there are POVM's that give probability distributions
that are equivalent to choose randomly among different
apparatuses.
\subsection{Group covariant POVM's}
Let's consider now the general scenario in which a group of
physical transformations $\gG$ can act on the probability space
${\Klm}$. We will write $gx$ for the action of the group
element $g\in\gG$ on the point $x\in{\Klm}$, and $g B$
for the  action of $g$ on a whole subset $B\subseteq\Klm$.
We will always consider the case in which $\gG$ acts 
transitively on $\Klm$, namely for any two points on 
$\Klm$ there is always a group element which connects them. 
A consequence of transitivity is that $\Klm$ can be always regarded as 
the homogeneous factor space $\Klm=\gG/\gG_x$,
$\gG_x$ denoting the stability group of any point $x\in\Klm$.
\par A POVM $P$ on $\sH$ for the probability space $\Klm$ is 
covariant under the unitary representation $g\to U_g$ of the group
$\gG$ when for every set  $B\in\sigma({\Klm})$ one has
\begin{equation}
U_g^\dag P(B)U_g=P(g^{-1}B).\label{covariantP}
\end{equation}
The following general theorem by Holevo\cite{Holevo82} classifies all
group-covariant POVM's.
\begin{theorem}[Holevo]\label{th:covdens} 
For square-integrable representations,
a POVM $P$ on the probability space $\Klm$ is covariant with 
respect to the unitary representation $g\to U_g$ on $\sH$ of the group
$\gG$ of transformations of  ${\Klm}$ if and only if it admits a density of
the form  
\begin{equation}
\d P_x =U_{g_x}^\dag \Xi U_{g_x}\d x,\quad g_x\in\gG\; :\,g_x x_0=x ,
\label{covdens}
\end{equation}
where $\d x$ is an invariant measure on $\Klm$, with $\Xi\ge 0$ in the
commutant $\gG_{x_0}'$ of the isotropy group $\gG_{x_0}$ of $x_0$, 
satisfying the constraint 
\begin{equation}
\int_\gG\d g\, U_g^\dag \Xi U_g=I_\sH,\label{constrdens}
\end{equation}
with $\d g$ invariant measure on $\gG$.
\end{theorem}
In the case in which the POVM is designed to estimate the group
element itself $g\in\gG$ corresponding to an unknown transformation
$U_g$, then the stability group is the identity, whence $\Klm=\gG$ and the  
POVM $P$ is covariant if and only if it admits a density of the form  
\begin{equation}
\d P_g =U_g^\dag\Xi U_g\d g,\quad g\in\gG\,
\label{covdens2}
\end{equation}
for any $\Xi\ge 0$ satisfying the constraint (\ref{constrdens}). The
possible {\em seed} operators $\Xi\ge 0$ satisfying the constraint 
(\ref{constrdens}) form a convex set. In Section \ref{extrcovP} we will
classify all extremal elements $\Xi$ of such convex set. 
\section{Quantum operations}\label{qo}
The mathematical structure that describes the most general state
change in quantum mechanics---such as the evolution of an open system
or the state change due to a measurement---is the {\em quantum
operation} (QO) of Kraus \cite{Kraus83a,Nielsen}. Such abstract theoretical evolution has
a precise physical counterpart in its implementations as a unitary
interaction between the system undergoing the QO and a part of the
apparatus---the so-called {\em ancilla}---which after the interaction
is read by means of a conventional quantum measurement.  
We can consider generally different input and output
Hilbert spaces $\sH$ and $\sK$, respectively, allowing the treatment
of very general quantum machines, e. g. of the kind of quantum optimal
cloners \cite{wern,clon}. For example in the cloning from one to $n$
copies one has input space $\sH$ and output space $\sK=\sH^{\otimes
n}$, or its symmetric version $\sK=\left(\sH^{\otimes n}\right)_+$ for symmetric cloning. 
Within the present paper we will only consider finite dimensional Hilbert spaces.
In the Heisenberg picture the QO
evolves observables, and will be denoted by a map $\map{M}$ from
$\Bndd{\sK}\rightarrow\Bndd{\sH}$. In the Schr\H{o}dinger picture the
QO evolves states, and it is given by the dual map
$\dual{\map{M}}:\T{\sH}\rightarrow\T{\sK}$, the dualism 
being determined by the equivalence of 
the two pictures in terms of the trace inner product, namely
$\Tr[\map{M}(X)\rho]=\Tr[\dual{\map{M}}(\rho)X]$ for all
$\rho\in\T{\sH}$ and for all $X\in\Bnd{K}$.
The maps $\map{M}$ and $\dual{\map{M}}$ are linear \emph{completely positive}
(CP),  namely they preserve positivity of the input operator for any
trivial extension $\map{M}\otimes\map{I}$ on a larger Hilbert space
that includes any possible additional quantum system, $\map{I}$ denoting the identity map on the
additional system. In the Schr\H{o}dinger picture the CP
property physically means that the map $\dual{\map{M}}$ from $\T{\sH}$
to $\T{\sK}$ preserves positivity of any input state of the quantum
system (with Hilbert space $\sH$) entangled with any possible
additional quantum system. The map $\dual{\map{M}}$ of a 
QO must also be trace-not-increasing, with the trace
$\Tr[\dual{\map{M}}(\rho)]\le 1$ representing the probability that the
transformation occurs, and the input and output states being connected 
as follows
\begin{equation}\label{start}
\rho\longmapsto\rho'=\frac{\dual{\map{M}}(\rho)}{\Tr[\dual{\map{M}}(\rho)]}.
\end{equation}
By denoting with $I_\sH$ the identity
operator on the Hilbert space $\sH$, we see that the
trace-not-increasing condition along with positivity of the map are
equivalent to the constraint
\begin{equation}
\map{M}(I_\sK)=K\in\Bndd{\sH},\qquad 0\le K\le I_\sH.\label{constrK}
\end{equation}
For finite-dimensional Hilbert spaces it is convenient to represent the maps  $\map{M}$ from
$\Bndd{\sK}\rightarrow\Bndd{\sH}$ as operators $R_\map{M}$ on
$\sK\otimes\sH$ using the following one-to-one correspondence
\begin{equation}
R_\map{M}=\dual{\map{M}}\otimes\map{I}(|I\>\< I|),\qquad
\dual{\map{M}}(\rho)=\Tr_\sH[(I_\sK\otimes\transp{\rho})R_\map{M}],\label{RMcorr}
\end{equation}
where $|I\>=\sum_n|n\>\otimes|n\>$ is a fixed vector in
$\sH\otimes\sH$, $\{|n\>\otimes|m\>\}$ denotes an orthonormal basis 
for $\sH\otimes\sH$, and the transposition $\tau$ for operators is defined with
respect to the orthonormal basis $|n\>\< m|$ for $\Bndd{\sH}$ taken as real.
One can easily check the correspondence (\ref{RMcorr}), and
injectivity follows from linearity. In addition, the operator
$R_\map{M}$ is non-negative if and only if the map $\map{M}$ is CP,
and the constraint (\ref{constrK}) in terms of the operator $K$
rewrites as follows
\begin{equation}
\Tr_\sK[R_\map{M}]=K,\qquad 0\le K\le I_\sH.\label{constrK2}
\end{equation}
The positive operators $R_\map{M}$ satisfying the
constraint (\ref{constrK2}) make a convex set, which is the operator
counterpart of the convex set of the corresponding QO's $\map{M}$.
\subsection{Group covariant CP-maps.}
We call the map $\map{M}$ from $\Bndd{\sK}$ to $\Bndd{\sH}$ $\gG$-covariant, when 
\begin{equation}
\map{M}(V_g^\dag X V_g)=U_g^\dag \map{M}(X)U_g,\quad\forall
g\in\gG,\label{covM}
\end{equation}
$\{U_g\}$ and $\{V_g\}$ denoting unitary representations of $\gG$
over the input and output spaces $\sH$ and $\sK$, respectively. The Schr\H{o}dinger picture
version of identity (\ref{covM}) is
\begin{equation}
\dual{\map{M}}(U_g\rho U_g^\dag)=V_g\dual{\map{M}}(\rho)V^{\dag}_g,\quad\forall
g\in\gG,\label{covMSc}
\end{equation}
where $\dual{\map{M}}$ goes from $\T{\sH}$ to  $\T{\sK}$.
\par The operator form $R_\map{M}$ for maps $\map{M}$ simplifies the
classification of QO's that are covariant under a group $\gG$,
resorting to the Wedderburn's decomposition of the commutant of the representation. 
It is easy to show that the map $\map{M}$ is $\gG$-covariant
(\ie it satisfies Eq. (\ref{covM})) if and only if its corresponding operator $R_{\map{M}}$ is
invariant under the representation $V_g\otimes U_g^*$\cite{clon}. In fact, from
Eq. (\ref{RMcorr}) using invariance of partial trace under cyclic
permutation of operators acting only on the traced space one has 
\begin{equation}
\begin{split}
0=&\dual{\map{M}}(\rho)-V_g^\dag\dual{\map{M}}(U_g\rho
U_g^\dag)V_g\\=&\Tr_\sH\{(I_\sK\otimes\transp{\rho})[R_\map{M}-(V_g^\dag\otimes
\transp{U_g})R_\map{M} (V_g\otimes U_g^*)]\},
\end{split}
\end{equation}
and, since Eq. (\ref{RMcorr}) is a one-to-one correspondence
between maps and operators, one concludes that
\begin{equation}
[R_{\map{M}},V_g\otimes U^*_g]=0,\quad\forall g\in\gG.
\end{equation}
Therefore, the problem of classifying covariant CP-maps resorts to
that of classifying positive elements of the commutant of the
representation  $V_g\otimes U^*_g$ on $\sK\otimes\sH$. 
By labeling with $k$ the generic equivalence class of the
representation, with multiplicity $m_k$, the Wedderburn's decomposition of the
representation space is written as follows\cite{Zhelobenko}
\begin{equation}
\sK\otimes\sH=\bigoplus_k (\sH_k\otimes \Cmplx^{m_k}).\label{Wedderburn}
\end{equation}
Then, since $R_\map{M}$ must be a positive operator in the commutant
of the representation it must have the general form
\begin{equation}
R_\map{M}=\oplus_k (I_{\sH_k}\otimes w_k^\dag w_k)= W^\dag W,\quad
W\doteq\oplus_k (I_{\sH_k}\otimes w_k),\label{Rcov0}
\end{equation}
where $w_k$ is any operator on $\Cmplx^{m_k}$, \ie a $m_k\times m_k$
matrix. Therefore, the classification of 
covariant trace-not-increasing QO's with
$\map{M}(I_\sK)=K\le I_\sH$ is equivalent to classify the operators
$R_\map{M}$ of the form (\ref{Rcov0}) with the constraint
\begin{equation}
\sum_k \Tr_\sK[(I_{\sH_k}\otimes w_k^\dag w_k)]=K\le I_\sH.\label{Kcov0}
\end{equation}
The constraint (\ref{Kcov0}) is generally quite involved, due to the
subspace mismatch between the tensor product 
$\sK\otimes\sH$ and the Wedderburn's decomposition: its simplification
will be the main task of Section \ref{excovqo}.
\section{Technical lemmas}\label{techlemma}
This section will be entirely devoted to technical lemmas, which will
be used for the classification of both extremal covariant POVM's and
QO's. The lemmas connect conditions on the vanishing of
partial traces with linear spannings.
\par In the following we will make use of the following simple fact 
for any linear space $\sL$ and a subspace
$\sS\subseteq\sL$: if the only vector of $\sL$ that is
orthogonal to the whole subspace $\sS$ is the null vector, then
one has $\sS=\sL$. Moreover, since orthogonality to a set
$\set{s}$ of vector implies orthogonality to its linear span
$\Span(\set{s})$, then the previous assertion holds also for subsets
$\set{s}\subseteq\sL$ (not necessarily subspace), namely if the
only vector orthogonal to the subset $\set{s}$ is the null vector,
than one has $\sL\equiv\Span(\set{s})$.  From now we will also
make use of the following natural notation
\begin{equation}
X(\Bndd{\sA}\otimes I_{\sB})Y^\dag\doteq\Span\{X(A\otimes
I_{\sB})Y^\dag, A\in\Bndd{\sA}\},\label{natnotat}  
\end{equation}
for $X,Y$ any operators with domain $\sA\otimes\sB$. 
\begin{lemma}\label{l:partr} Let $B\in\Bndd{\sB_2\otimes\sB_1,\sA}$, $\sA$ and
$\sB_{1,2}$ denoting arbitrary finite dimensional Hilbert spaces. 
Then, the injectivity of the linear CP
map $\map{W}(A)=\Tr_{\sB_1}[B^\dag A B]$ on $\Bnd{A}$ is equivalent to
the spanning condition
\begin{equation}
\Bnd{A}= B(\Bndd{\sB_2}\otimes I_{\sB_1})B^\dag.\label{iff2}
\end{equation}
\end{lemma}
\Proof
The iniectivity of the map $\map{W}(A)=\Tr_{\sB_1}[B^\dag A B]$ on
$\Bnd{A}$ means that
\begin{equation}
\forall A\in\Bnd{A}\quad\Tr_{\sB_1}[B^\dag A B]=0\Longrightarrow A=0.\label{iff1}
\end{equation}
The condition $\Tr_{\sB_1}[B^\dag A B]=0$ is equivalent to 
$\Tr[C\Tr_{\sB_1}[B^\dag A B]]=0$  $\forall C\in\Bndd{\sB_2}$. Therefore,
since one has 
\begin{equation}
\Tr[C\Tr_{\sB_1}[B^\dag A B]]=\Tr[(C\otimes I_{\sB_1})B^\dag
A B]=\Tr[B(C\otimes I_{\sB_1})B^\dag A]
\end{equation}
condition (\ref{iff1}) is then equivalent to
\begin{equation}
\forall A\in\Bnd{A},\; \Tr[B(\Bndd{\sB_2}\otimes I_{\sB_1})B^\dag
A]=0\;\Longrightarrow A=0,\label{iff3}
\end{equation}
where we used notation (\ref{natnotat}). Eq. (\ref{iff3}) says that
the only operator $A\in\Bnd{A}$ orthogonal to the operator space 
$B(\Bndd{\sB_2}\otimes I_{\sB_1})B^\dag\subseteq\Bnd{A}$ is the null operator, which
means that $B(\Bndd{\sB_2}\otimes I_{\sB_1})B^\dag$ is actually the
full linear space $\Bnd{A}$, namely condition (\ref{iff3}) is
equivalent to condition (\ref{iff2}).\qed
\bigskip
The above theorem leads immediately to the following corollaries. 
\begin{corollary} A necessary condition for injectivity of the map
$\map{W}(A)=\Tr_{\sB_1}[B^\dag A B]$ on $\Bnd{A}$ is
\begin{equation}
\dim(\sA)\le\min\{\dim(\sB_2),\rnk(B)\}.
\end{equation}
\end{corollary}
\bigskip
\begin{corollary}
The injectivity of the map $\map{W}(A)=\Tr_{\sB_1}[B^\dag A B]$ on
$\Bnd{A}$ is equivalent to the existence of a linear
injective map $\map{V}$ from $\Bnd{A}$ to $\Bnd{B_2}$ such that
\begin{equation}
\forall A\in\Bnd{A}\quad B(\map{V}(A)\otimes
I_{\sB_1}) B^\dag=A.\label{mapV}
\end{equation}
The relation between the maps $\map{W}$ and $\map{V}$ is given by
\begin{equation}
\map{W}(A)=\Tr_{\sB_1}[B^\dag B(\map{V}(A)\otimes
I_{\sB_1}) B^\dag B].\label{WV}
\end{equation}
\end{corollary}
\Proof The spanning condition (\ref{iff2})---equivalent to the injectivity of the
map $\map{W}(A)=\Tr_{\sB_1}[B^\dag A B]$ on
$\Bnd{A}$---guarantees that for each $A\in\Bnd{A}$
there exists an element, say $V_A$, of $\Bnd{B_2}$ such that $B(V_A\otimes
I_{\sB_1}) B^\dag=A$. Consider now an orthonormal basis $A_j$ for
$\Bnd{A}$, and denote by  $V_j$ any element of  of
$\Bnd{B_2}$ such that $B(V_j\otimes I_{\sB_1})
B^\dag=A_j$. It is clear that the $\{V_j\}$ can be chosen as linearly
independent. Now, for every element
$A\in\Bnd{A}$ define $\map{V}(A)=\sum_j\Tr[A_j^\dag
A]V_j$. This map is clearly linear and injective. The map
$\map{V}(A)$ corresponds to a nonorthogonal change of basis (from
$\{A_j\}$ to $\{V_j\}$) which compensates the nonorthogonal change
of basis $B(V_j\otimes I_{\sB_1}) B^\dag=A_j$. Eq. (\ref{WV}) 
follows by substituting Eq. (\ref{mapV}) into the map $\map{W}$.\qed
\bigskip
We have also the additional lemma.
\begin{lemma}\label{l:partr2} As in Lemma \ref{l:partr}, the 
injectivity of the map $\map{W}(A)=\Tr_{\sB_1}[B^\dag A B]$ on
$\Bnd{A}$ is equivalent to the linear independence of
the set of operators $\{W_i^\dag W_j\}$, where
$W_i\in\Bndd{\sB_1,\sB_2}$ are defined from the singular value
decomposition $B=\sum_i|V_i\>\< W_i|$ through the identity  
$|W_i\>=(W_i\otimes I_{\sB_1})|I\>$, $|I\>\in\sB_1^{\otimes 2}$
denoting the fixed vector $|I\>=\sum_l|l\>\otimes|l\>$, for $\{|l\>\otimes|m\>\}$
arbitrary orthonormal basis of $\sB_1^{\otimes2}$.
\end{lemma}
\Proof
\par First, notice that the identity $|X\>=(X\otimes I_{\sB_1})|I\>$
sets a bijection between vectors $|X\>\in\sB_2\otimes\sB_1$
and operators $X\in\Bndd{\sB_1,\sB_2}$. 
Then, using the singular value decomposition $B=\sum_i|V_i\>\< W_i|$,
with $|V_i\>\in\sA$ and $|W_i\>\in\sB_2\otimes\sB_1$, the partial
trace in Eq. (\ref{iff1}) becomes
\begin{equation}
\Tr_{\sB_1}[B^\dag A B]=\sum_{ij}\< V_i|A|V_j\>\Tr_{\sB_1}[|W_i\>\<W_j|]=
\sum_{ij}\< V_i|A|V_j\>\transp{W}_i W_j^*,
\end{equation}
where $\tau$ denotes the transposition for which 
$(X\otimes I_{\sB_1})|I\>=(I_{\sB_1}\otimes\transp{X})|I\>$, and $*$
denotes complex conjugation, i. e. $X^\dag=(\transp{X})^*$.
By taking the complex conjugate of the last equation and
introducing the matrix 
$A_{ij}\doteq\< V_i|A|V_j\>^*\in\set{M}_N(\Cmplx)$ where $N=\rnk(B)$
($N^2$ is the cardinality of the set $\{W_i^\dag W_j\}$),  the statement  
(\ref{iff1}) is equivalent to
\begin{equation}
\{A_{ij}\}\in\set{M}_N(\Cmplx),\quad\sum_{ij}A_{ij}W_i^\dag
W_j=0\;\Longrightarrow A_{ij}=0,\;\forall\, i,j,\label{Aij} 
\end{equation}
namely the operators $\{W_i^\dag W_j\}$ are linearly independent.\qed
\bigskip
\par In the following we will need the following generalization of Lemma \ref{l:partr}.
\begin{lemma}\label{l:partr3} Let
  $B\in\Bndd{\oplus_k(\sB_2^{(k)}\otimes\sB_1^{(k)}),\sA}$, and denote
  by $P_k$ the orthogonal projector over $\sB_2^{(k)}\otimes\sB_1^{(k)}$,
$\sA$ and $\sB_{1,2}^{(k)}$ being arbitrary finite dimensional Hilbert spaces.
\par The following implication
\begin{equation}
A\in\Bnd{A},\Tr_{\sB_2^{(k)}}[P_kB^\dag A BP_k]=0\,\forall
k\Longrightarrow A=0.\label{iff1k}
\end{equation}
is equivalent to
\begin{equation}
\Bnd{A}=\Span\{
B[\oplus_k(\Bndd{\sB_2^{(k)}}\otimes I_{\sB_1^{(k)}})]B^\dag\},\label{iffk}
\end{equation}
and necessary conditions are
\begin{eqnarray}
\dim(\sA)^2&\le&\sum_k\dim(\sB_2^{(k)})^2,\\
\dim(\sA)&\le&\rnk(B).
\end{eqnarray}
\end{lemma}
\Proof
The condition $\Tr_{\sB_1^{(k)}}[P_kB^\dag A BP_k]=0$ $\forall k$ is equivalent to
say that for any $C_k\in\Bndd{\sB_2^{(k)}}$ one has
$\Tr[P_kC_k\Tr_{\sB_1^{(k)}}[P_kB^\dag A BP_k]]=0$ $\forall k$. Since
one has
\begin{equation}
\begin{split}
\Tr[C_k\Tr_{\sB_1^{(k)}}[P_kB^\dag A BP_k]]&=\Tr[(C_k\otimes I_{\sB_1^{(k)}})P_kB^\dag
A BP_k]\\&=\Tr[BP_k(C_k\otimes I_{\sB_1^{(k)}})P_kB^\dag A],
\end{split}
\end{equation}
and, therefore, condition (\ref{iff1k}) is equivalent to
\begin{equation}
A\in\Bnd{A},\; \Tr[BP_k(\Bndd{\sB_2^{(k)}}\otimes I_{\sB_1^{(k)}})P_kB^\dag
A]=0\,\forall k\;\Longrightarrow A=0.\label{iff3k}
\end{equation}
The last condition says that the only operator in $\Bnd{A}$ which is
orthogonal to the set $BP_k(\Bndd{\sB_2^{(k)}}\otimes
I_{\sB_1^{(k)}})P_kB^\dag$ $\,\forall k$ is the null operator, or, in
other words that the set spans the full operator space $\Bnd{A}$,
namely Eq. (\ref{iffk}). The necessary conditions then follow trivially.
\qed
\bigskip
\par We are now ready to classify the extremal group
covariant POVM's and QO's in the following sections. 
In order to classify extremal elements of convex sets, we will use the
method of perturbations. We will call a non null operator $B$ a {\em perturbation} for an operator
$A$ in a convex set if both $A\pm tB$ are still in the convex set for some (sufficiently
small) $t>0$. Then, clearly $A$ is not extremal in the convex set if
and only if it has a perturbation.
\section{Extremal covariant POVM's}\label{extrcovP}
We have seen that the covariant POVM for the estimation of a group element $g$  
of an unknown unitary transformation $U_g$ is of the general form
\begin{equation}
\d P_g=\d g\,U_g^\dag \Xi U_g^\dag,
\end{equation}
with probability space $\Klm=\gG$, and with 
\begin{equation}
\int_\gG\d g\, U_g^\dag \Xi U_g=I_\sH.\label{constr2}
\end{equation}
The Wedderburn's decomposition (\ref{Wedderburn}) of the
representation space here rewrites as follows
\begin{equation}
\sH=\bigoplus_k (\sH_k\otimes \Cmplx^{m_k}),\label{Wedderburn2}
\end{equation}
where we remind that $k$ labels the equivalence class of
irreducible components, and $m_k$ denotes its multiplicity.
The integral in the normalization condition (\ref{constr2})
belongs to the commutant of the representation, whence it can be
rewritten as follows
\begin{equation}
\int_\gG\d g\, U_g^\dag \Xi U_g=\bigoplus_k d_{\sH_k}^{-1}\Big[I_{\sH_k}\otimes
\Tr_{\sH_k}(P_k\Xi P_k)\Big]=I_\sH,\label{Wedderburn3}
\end{equation}
$P_k$ denoting the orthogonal projector on the subspace
$\sH_k\otimes \Cmplx^{m_k}$. Eq. (\ref{Wedderburn3}) follows from the
simple fact that for an irreducible representation on the space say
$\sL$, one has $\int_\gG\d g\, U_g^\dag Z U_g=d_\sL^{-1}
\Tr[Z]I_\sL$ for measure $\d g$ normalized to unit on $\gG$.
Eq. (\ref{Wedderburn3}) allows to split the constraint
(\ref{constr2}) into the following set of constraints
\begin{equation}
\Tr_{\sH_k}(P_k\Xi P_k)=d_{\sH_k} I_{m_k},\;\forall k,\label{POVMconstrG}
\end{equation}
where by $I_{m_k}$ we denote the identity matrix over $\Cmplx^{m_k}$.
We then conclude that the classification of extremal $\gG$-covariant POVM's is
equivalent to find the extremal $\Xi$ within the convex set
of operators $\Xi\ge 0$ satisfying the constraints
(\ref{POVMconstrG}).  For such purpose we have the following theorem.
\begin{theorem}\label{t:covPOVM}
Let $\Xi$ be an element of the convex set of positive operators on
$\sH$ satisfying the constraints
\begin{equation}
\Tr_{\sH_k}(P_k\Xi P_k)=d_{\sH_k} I_{m_k},\qquad\forall k\in\set{S},
\end{equation}
where $\set{S}$ denotes the set of equivalence classes of irreducible
components in the representation.
Write $\Xi$ in the form $\Xi=X^\dag AX$ with  $A\ge 0$, choosing $\Rng(X)=\Supp(A)
\doteq\Ker(A)^\perp$. Then
\par {\bf 1.} $\Theta$ is a perturbation of $\Xi$ if and only if
$\Theta$ is Hermitian, with $\Tr_{\sH_k}(P_k\Theta P_k)=0$ $\forall k\in\set{S}$, and
$\Theta=X^\dag BX$ for some nonzero Hermitian $B$ with 
$\Supp(B)\subseteq\Supp(A)$. 
\par {\bf 2.} For the specific choice of the form of $A$ as
$A=\oplus_k A_k$, with $A_k\in\Bndd{\sH_k\otimes\Cmplx^{m_k}}$, one has $B=\oplus_k
B_k$, $B_k\in\Bndd{\sH_k\otimes\Cmplx^{m_k}}$ and
$\Supp(B_k)\subseteq\Supp(A_k)$,  $\forall k\in\set{S}$; 
\par {\bf 3.} $\Xi=X^\dag X$ is extremal if and only if
\begin{equation}
\Bndd{\Rng(X)}=\Span\{X[\oplus_k(I_{\sH^{(k)}}\otimes\Bndd{\Cmplx^{m_k}})]X^\dag\}.
\label{newiff}
\end{equation}
\end{theorem}
\Proof 
\par {\bf 1.} Let $\Theta$ Hermitian, with $\Tr_{\sH_k}(P_k\Theta P_k)=0$, and
$\Theta=X^\dag BX$ for some nonzero Hermitian $B\in\Bndd{\sH}$ and with
$\Supp(B)\subseteq\Supp(A)$. Then for $\rnk(B)>0$ $\Theta$ is
necessarily nonzero, and since $A\ge 0$, both constraints $A\pm tB\ge
0$ and $\Tr_{\sH_k}(P_k(\Xi\pm t\Theta) P_k)=d_{\sH_k} I_{m_k}\;\forall k$ are satisfied for some
$t>0$, whence $\Theta$ is a perturbation for $\Xi$. Conversely,
suppose $\Theta\in\Bndd{\sH}$ is a perturbation for  $\Xi$. Since we must
have $\Xi\pm t\Theta\ge 0$ and $\Tr_{\sH_k}[P_k(\Xi\pm
t\Theta)P_k]=d_{\sH_k} I_{m_k}$ for some $t>0$, then $\Theta$ is Hermitian
with $\Tr_{\sH_k}(P_k\Theta P_k)=0$  $\forall k\in\set{S}$. Moreover, if we write $\Xi$ in the
form $\Xi=X^\dag AX$ with nonnegative $A\in\Bndd{\sH}$, and $\Rng(X)=\Supp(A)$, then
also $\Theta$ can be written in the same form $\Theta=X^\dag BX$ for some
nonzero Hermitian $B\in\Bndd{\sH}$ and $\Tr_{\sH_k}[P_k(\Xi\pm
t\Theta)P_k]=d_{\sH_k} I_{m_k}$.  
In fact, if $X$ is not invertible, it can be always completed to an 
invertible operator $Z=X+Y$ by adding an operator $Y$ with
$\Rng(Y)=\Ker(A)$, and one can equivalently write $\Xi=Z^\dag AZ$. Now
we can write also the perturbation operator in the form
$\Theta=Z^\dag BZ$. However, since $A\pm t B\ge 0$ for
some $t$, then necessarily $B$ must have $\Supp(B)\subseteq\Supp(A)=\Rng(X)$,
whence $Z^\dag BZ=X^\dag BX$. 
\par {\bf 2.} First it is obvious that a choice of the form
$A=\oplus_k A_k$, with $A_k\in\Bndd{\sH_k\otimes\Cmplx^{m_k}}$ is
always possible. Then, in order to have $A\pm t B\ge 0$ for some $t>0$, one must have 
$B=\oplus_k B_k$, each $B_k$ Hermitian, with
$\Supp(B_k)\subseteq\Supp(A_k)$,  $\forall k\in\set{S}$.
\par {\bf 3.} Since $\Supp(A)\subseteq\Rng(X)$ and $A\ge 0$, we can
always merge $\sqrt{A}$ into $X$ by substituting $X\to \sqrt{A}X $.
Then, since $\Xi$ is not extremal iff it has a
perturbation, by part {\bf 1} one sees that $\Xi$ is extremal iff
for Hermitian $B\in\Bnd{H}$ with $\Supp(B)\subseteq\Rng(X)$, one has
\begin{equation}
\Tr_{\sH_k}(P_k X^\dag BX
P_k)=0\,\forall k\in\set{S}\quad\Longrightarrow\quad B=0, 
\end{equation}
whence via Cartesian decomposition of $B$ we have the equivalent statement
\begin{equation}
B\in\Bndd{\Rng(X)},\;\Tr_{\sH_k}(P_k X^\dag BX
P_k)=0\,\forall k\in\set{S}\quad\Longrightarrow\quad B=0. 
\end{equation}
Then, by Lemma \ref{l:partr3} this is equivalent to condition
(\ref{newiff}).\qed 
\begin{corollary}
A necessary condition for extremality of the seed $\Xi$ of a group
covariant representation as in Theorem \ref{t:covPOVM} is
\begin{equation}
\rnk(\Xi)^2\le\sum_k m_k^2.\label{POVnec2}
\end{equation}
\end{corollary}
\Proof Eq. (\ref{POVnec2}) is a trivial consequence of the necessary condition (\ref{newiff}).\qed
\begin{corollary}
Every rank-one POVM is extremal.
\end{corollary}
\Proof For $\rnk(X)=1$ the iff condition (\ref{newiff}) is trivially satisfied.\qed 
\begin{theorem}\label{t:onek}
For $\set{S}$ containing only a single equivalence class, say $h$,
with multiplicity $m_h\ge1$, the extremality of a covariant POVM on the
Hilbert space $\sH=\sH_h\otimes\Cmplx^{m_h}$ is equivalent to the
linear independence of the set of operators $\{W_i^\dag W_j\}$, where
$W_i\in\Bndd{\Cmplx^{m_h},\sH_h}$ are defined from the
spectral decomposition $\Xi=\sum_i|W_i\>\< W_i|$ of the seed $\Xi$ of
the POVM through the identity $|W_i\>=(W_i\otimes I_{m_h})|I\>$,
$|I\>\in(\Cmplx^{m_h})^{\otimes2}$ denoting the fixed vector
$|I\>=\sum_l|l\>\otimes|l\>$, for $\{|l\>\otimes|m\>\}$ 
arbitrary orthonormal basis of $(\Cmplx^{m_h})^{\otimes2}$. Extremal
POVM's with any rank $\rnk(\Xi)\le m_h$ are admissible.
\end{theorem}
\Proof For $\set{S}$ containing a single equivalence class $h$ with
multiplicity $m_h\ge1$ the seed $\Xi$ of the POVM 
must satisfy the single constraint
\begin{equation}
\Tr_{\sH_h}(\Xi)=d_{\sH_h} I_{m_h}. 
\end{equation}
Now, write $\Xi$ in the form $\Xi=X^\dag AX$ with
$X\in\Bndd{\sH_h\otimes\Cmplx^{m_h},\sA}$, and  
$\Rng(X)=\Supp(A)$, $\sA$ being a Hilbert space such that
$\Supp(A)\subseteq\sA\subseteq\sH_h\otimes\Cmplx^{m_h}$, and which
can be chosen as $\sA\simeq\Rng(X)$. Then, according to Theorem \ref{t:covPOVM} 
$\Theta$ is a perturbation for $\Xi$ iff
it is of the form $\Theta=X^\dag BX$, with $B$ Hermitian, 
$\Supp(B)\subseteq\Supp(A)$, and $\Tr_{\sH_h}(X^\dag
BX)=0$. This means that the extremality of $\Xi$ is equivalent to the
injectivity of the map $\map{W}(B)=\Tr_{\sH_h}(X^\dag BX)$ over the set
of Hermitian operators $B$ with $\Supp(B)\subseteq\Supp(A)$, which is
equivalent to injectivity of the same map on $\Bndd{\Rng(X)}$. We are
thus in the situation of Lemma \ref{l:partr2}, with
$\sA=\Rng(X)$, $\sB_1=\Cmplx^{m_h}$ and $\sB_2=\sH_h$.
Therefore, by writing the singular value decomposition of
$X=\sum_i|V_i\>\< W_i|$, with $\Span\{|V_i\>\}=\Rng(X)=\Supp(A)$
the injectivity of the map $\map{W}(B)=\Tr_{\sH_h}[X^\dag B X]$ on
$\Bnd{\Rng(X)}$ is equivalent to the linear independence of
the set of operators $\{W_i^\dag W_j\}$, where
$W_i\in\Bndd{\Cmplx^{m_h},\sH_h}$ are defined through the identity  
$|W_i\>=(W_i\otimes I_{m_h})|I\>$, $|I\>\in(\Cmplx^{m_h})^{\otimes 2}$
denoting the fixed vector $|I\>=\sum_l|l\>\otimes|l\>$, with 
$\{|l\>\otimes|m\>\}$ arbitrary orthonormal basis of
$(\Cmplx^{m_h})^{\otimes2}$. Now, the maximum rank of the POVM is
given by the maximum number of operators $W_i$ such that the set of
operators $\{W_i^\dag W_j\}$ in $\Bndd{\Cmplx^{m_h}}$ is linearly
independent. Since we can have at most $m_h^2$  linearly
independent operators in $\Bndd{\Cmplx^{m_h}}$, the maximum cardinality of the
set $\{W_i\}$ is $m_h$.\qed
\begin{corollary}
A POVM which is covariant  under an irreducible representation is extremal iff it
is rank one.
\end{corollary}
\Proof For $\set{S}$ containing a single equivalence class $h$ with
multiplicity $m_h=1$ the iff condition (\ref{newiff}) rewrites
\begin{equation}
\Bndd{\Rng(X)}= \Span\{X (I_{\sH^{(h)}}\otimes\Cmplx^1) X^\dag\}=\Span\{XX^\dag\},
\end{equation}
which is satisfied iff $\rnk(X)=1$. As an alternative proof, the
present corollary corresponds to the situation of Theorem \ref{t:onek}
for multiplicity $m_h=1$.\qed 
\bigskip
\example Consider a POVM on $\sH$ with $\dim(\sH)=d$ covariant under
$\gG=\U{1}$, with 
\begin{equation}
U_\phi=\exp(i\phi N),\qquad N=\sum_{n=0}^{d-1}n|n\>\< n|.
\end{equation}
Here we have $d$ one-dimensional irreducible representations with
characters $\chi_k(\phi)=\exp(ik\phi)$, $k=0,\ldots d-1$, namely they
are all inequivalent, whence with unit multiplicity. Therefore,
the necessary condition (\ref{POVnec2}) bounds the rank of the POVM as
follows
\begin{equation}
\rnk(\Xi)^2\le\dim(\sH),
\end{equation}
and in order to have $\rnk(\Xi)=2$ one must have $\dim(\sH)\ge 4$.
According to Theorem \ref{t:covPOVM} the extremal POVM's have seed of the form $\Xi=X^\dag X$
satisfying the identity 
\begin{equation}
\Bndd{\Rng(X)}=\Span\{|X_k\>\< X_k|:\; 0\le k\le\dim(\sH)\}.
\end{equation}
where $|X_k\>=X|k\>$, $\{|k\>\}$ denoting any orthonormal basis for
$\sH$. Notice that in the present example the operator 
$\Xi$ corresponds to a so-called {\em correlation matrix},
namely a positive matrix with all ones on the diagonal. This follows
from the constraint  (\ref{POVMconstrG}), which in our case is simply
$\<k|\Xi|k\>=1,\,\forall\, k$. Therefore, the present classification
of extremal POVM's coincides with the classification of extremal
correlation matrices given in Ref. \cite{LiTam94}.
\example Consider a POVM for $n$ qubits on the Hilbert space
$\sH=(\Cmplx^2)^{\otimes n}$ covariant under the tensor representation
$U_\phi^{\otimes n}$ of $\gG=\U{1}$, with  
\begin{equation}
U_\phi=\exp(i\phi |1\>\< 1|),
\end{equation}
where $\{|0\>,|1\>\}$ is a orthonormal basis for $\Cmplx^2$.
Here we have $n+1$ one-dimensional irreducible representations with
characters $\chi_k(\phi)=\exp(ik\phi)$, $k=0,\ldots n$, and with
multiplicity $m_k={n\choose k}$. An orthonormal basis of each
subspace $\Cmplx^{m_k}$ of $\sH=\oplus_k \Cmplx^{m_k}$ is
given by 
\begin{equation}
\{|j\>_k\}=\{P_j^{(n,k)}\ket{\underbrace{00\ldots0}_{n-k}\underbrace{111\ldots1}_k}\}, 
\end{equation}
where $P_j^{(n,k)}$ denotes the $j$th permutation of $k$ qubits in the
state $|1\>$ in the tensor product of $n$ qubits in total, with
all other qubits  in the state $|0\>$. In the present example, 
the iff condition for extremality (\ref{newiff}) requires that
$\Xi=X^\dag X$ satisfies the identity 
\begin{equation}
\Bndd{\Rng(X)}= \Span\{X |i\>_k {}_k\< j| X^\dag,\, k\in\set{S}, i,j=1,\ldots m_k\},
\end{equation}
where now $\{|i\>_k\}$ denotes any orthonormal basis for
$\Cmplx^{m_k}$.  The necessary condition (\ref{POVnec2}) bounds the
rank of the POVM as follows
\begin{equation}
\rnk(\Xi)^2\le\sum_{k=0}^n {n\choose k}^2={2n\choose n}.
\end{equation}
Here, in order to have $\rnk(\Xi)\ge 2$ one needs $n\ge 2$ qubits. For
$n=2$ according to the previous example, one necessarily must have at
least two inequivalent classes, since each of the irreducible
components has less than four dimensions (the same 
is true also for $n=3$). The previous example is also recovered by
considering the special case in which
$\Rng(X)\subseteq((\Cmplx^2)^{\otimes n})_+$ \ie containing only the
sub-representation of $U_\phi^{\otimes n}$ on the symmetric subspace
$((\Cmplx^2)^{\otimes n})_+$, with multiplicity 1.
\example Consider a POVM on $\sH^{\otimes 2}$ which is covariant under
the group representation $U_g\otimes I_\sH$, where $U_g$ is an
irreducible representation of $\gG$ on $\sH$. Here, we trivially have
a single equivalence class, say $h$,  (corresponding to the irreducible 
representation $U_g$) with multiplicity $m_h=\dim(\sH)$, \ie the
Hilbert space $\sH$ coincides with the multiplicity space $\sH\simeq\Cmplx^{m_h}$.
This is exactly the case considered in Theorem
\ref{t:onek}. Therefore, the extremality of the POVM is equivalent to
the linear independence of the set of operators $\{W_i^\dag W_j\}$, where
$W_i\in\Bnd{H}$ are defined from the spectral decomposition
$\Xi=\sum_i|W_i\>\< W_i|$ of the seed $\Xi$ of the POVM through the
identity $|W_i\>=(W_i\otimes I_\sH)|I\>$,  as in Theorem \ref{t:onek}.
Therefore, we can have extremal POVM's with any 
$\rnk(\Xi)\le\dim(\sH)$. Notice that there cannot be more than a single
maximally entangled vector $|W_i\>$ in the decomposition of $\Xi$,
since, otherwise, at least two operators $W_i$ would be proportional 
to unitary operators, and then the set $\{W_i^\dag W_j\}$ would be
necessarily linearly dependent (two products would be both proportional to the
identity). The rank-one case with a single maximally entangled
projector corresponds to a so-called {\em Bell POVM}.
\section{Extremal covariant quantum operations}\label{excovqo}
In the following we will denote shortly by $\aA_\gG$ the operator
algebra generated by the group representation $V_g\otimes U_g^*$, by
$\aA_\gG'$ its commutant, and finally by $\aH_\gG'$ the Hermitian
operators in the commutant. The following theorem classifies all
extremal $\gG$-covariant maps $\map{M}$ in the convex set given by
Eq. (\ref{Kcov0}).
\begin{theorem}\label{th:extremalR}
Let $R$ be an element of the convex set of positive operators in the commutant
$\aA_\gG'$ of the operator algebra $\aA_\gG$ generated by the group
representation $V_g\otimes U_g^*$ on $\sK\otimes\sH$, \ie  of the form 
\begin{equation}
R=\oplus_k (I_{\sH_k}\otimes w_k^\dag w_k)=W^\dag W,\quad
W\doteq\oplus_k (I_{\sH_k}\otimes w_k),\label{Rcov}
\end{equation}
satisfying the constraint
\begin{equation}
\sum_k \Tr_\sK[(I_{\sH_k}\otimes w_k^\dag w_k)]=K\le I_\sH,\label{Kcov}
\end{equation}
where 
\begin{equation}
\sH\otimes\sK=\bigoplus_k (\sH_k\otimes\Cmplx^{m_k})
\end{equation}
is the Wedderburn's decomposition of the representation space, $k$
labeling the equivalence class of representations, with multiplicity
$m_k$. Denote by $P_k$ the orthogonal projector over the space
$\sH_k\otimes\Cmplx^{m_k}$ of the equivalence class.
Write $R$ in the form $R=X^\dag QX$ with $Q,X\in\aA_\gG'$
and $\Rng(X)=\Supp(Q)$. Then:
\par {\bf 1.} $S$ is a perturbation of $R$ if and only if $S\in\aH_\gG'$, with
$\Tr_\sK[S]=0$, and $S=X^\dag OX$ for some nonzero $O\in\aH_\gG'$ with 
$\Supp(O)\subseteq\Rng(X)$. Specifically, writing
$Q=\oplus_k(I_{\sH_k}\otimes Q_k)$ and $X=\oplus_k(I_{\sH_k}\otimes X_k)$, one has
$O=\oplus_k(I_{\sH_k}\otimes O_k)$ with $\Supp(O_k)\subseteq\Rng(X_k)$ $\forall\, k$.
\par {\bf 2.} One can always write $R$ in the form $R=X^\dag X$, with
$X\in\aA_\gG'$ of the form $X=\oplus_k(I_{\sH_k}\otimes X_k)$.
Denote by $\set{S}$ the set of equivalence classes $k$ for which
$X_k\neq 0$. Then, a necessary and sufficient condition for
extremality of $R=X^\dag X$ with $\Tr_\sK[R]=K$ is the injectivity of
the map $\map{T}(O)=\Tr_\sK[X^\dag OX]$ on $\aA_\gG'\cap\Bndd{\Rng(X)}$, namely
\begin{equation}
O\in\aA_\gG'\cap\Bndd{\Rng(X)},\; \Tr_\sK[X^\dag OX]=0\Longrightarrow
O=0,\label{iffmap}
\end{equation}
which is equivalent to 
\begin{equation}
\oplus_{k\in\set{S}}\Bndd{\Rng(X_k)}=\oplus_{k\in\set{S}}
X_k\Tr_{\sH_k}[P_k(I_\sK\otimes\Bnd{H})P_k]X_k^\dag.
\label{iffmap2}\end{equation}
\end{theorem}
\Proof 
\par {\bf 1.} Let $S\in\aH_\gG'$, with $\Tr_\sK[S]=0$, and $S=X^\dag OX$ for
some nonzero Hermitian $O$ with $\Supp(O)\subseteq\Supp(Q)$. Then for
$\rnk(O)>0$ $S\in\aH_\gG'$ is necessarily nonzero, and
since $\aH_\gG'\ni Q\ge 0$, all constraints: $Q\pm tO\in\aH_\gG'$,
$Q\pm tO\ge 0$, and $\Tr_\sK[R\pm tS]=K$ 
are satisfied for some $t>0$, whence $S$ is a perturbation for
$R$. Conversely, suppose that $S\in \sK\otimes\sH$ is a perturbation for 
$R$. Since we must have $\aH_\gG'\ni R\pm tS\ge 0$ and $\Tr_\sK[R\pm tS]=K$  
for some $t>0$, then $S\in\aH_\gG'$ with $\Tr_\sK[S]=0$. Moreover, if
we write $R$ in the form $R=X^\dag QX$ with $\Rng(X)=\Supp(Q)$, 
then also $S$ can be written in the form $S=X^\dag OX$ for some nonzero
Hermitian $O\in\aH_\gG'$.
In fact, if $X$ is not invertible, it can be always completed to an
invertible operator $Z=X+Y$ by adding an operator $Y\in\aA_\gG'$
of the form $Y=\oplus_k (I_{\sH_k}\otimes Y_k)$ with
$\Rng(Y_k)=\Ker(Q_k)$ (where $Q=\oplus_k (I_{\sH_k}\otimes Q_k)$), and one can
equivalently write $R=Z^\dag QZ$ with $Q\in\aH_\gG'$ and $Z\in\aA_\gG'$. Now 
we can write also the perturbation operator in the form
$S=Z^\dag OZ$. However, since for some $t$ the operator $Q\pm tO\ge 0$ must belong
to the commutant $\aA_\gG'$, then necessarily $O\in\aH_\gG'$ and
$\Supp(O)\subseteq\Supp(Q)=\Rng(X)$, with $Z^\dag OZ=
X^\dag OX$. Specifically, writing
$Q=\oplus_k(I_{\sH_k}\otimes Q_k)$, one has
$O=\oplus_k(I_{\sH_k}\otimes O_k)$
with $\Supp(O_k)\subseteq\Supp(Q_k)=\Rng(X_k)$ $\forall\, k$.
\par {\bf 2.} As in part {\bf 1} we can always take $Q$ as the
identity, and redefine $X\to\sqrt{Q}X$, since $Q\ge 0$,
keeping $X$ of the form $X=\oplus_k(I_{\sH_k}\otimes X_k)$, since both
operators in the product $\sqrt{Q}X$ belong to the algebra $\aA_\gG'$. 
From part {\bf 1} we then see that $R=X^\dag X$ with $X\in\aA_\gG'$ is extremal if and only if 
\begin{equation}
O\in\aH_\gG'\cap\Bndd{\Rng(X)},\; \Tr_\sK[X^\dag OX]=0\Longrightarrow O=0,
\end{equation}
and via Cartesian decomposition this is equivalent to
\begin{equation}
O\in\aA_\gG'\cap\Bndd{\Rng(X)},\; \Tr_\sK[X^\dag OX]=0\Longrightarrow
O=0.\label{Ocart}
\end{equation}
Since $O\in\aA_\gG'\cap\Bndd{\Rng(X)}$ can be decomposed as $O=\oplus_k
(I_{\sH_k}\otimes O_k)$ with $O_k\in\Bndd{\Rng(X_k)}$ $\forall
k\in\set{S}$, then the statement (\ref{Ocart}) is equivalent to 
\begin{equation}
\begin{split}
\forall k\in\set{S}&\;O_k\in\Bndd{\Rng(X_k)}, \\ 
\sum_{k\in\set{S}}&\Tr_\sK[(I_{\sH_k}\otimes X_k)^\dag 
 (I_{\sH_k}\otimes O_k) (I_{\sH_k}\otimes X_k)]=0\Longrightarrow
 O_k=0\,\forall k\in\set{S},
\end{split}
\end{equation}
or else
\begin{equation}
\begin{split}
\forall k\in\set{S}&\;O_k\in\Bndd{\Rng(X_k)}, \\ 
&\Tr_\sK[\oplus_{k\in\set{S}}(I_{\sH_k}\otimes X_k)^\dag 
 (I_{\sH_k}\otimes O_k) (I_{\sH_k}\otimes X_k)]=0\Longrightarrow
 O_k=0\,\forall k\in\set{S},\label{Ocart2}
\end{split}
\end{equation}
The vanishing of the partial trace can be written as the vanishing of the trace
$\Tr[\oplus_{k\in\set{S}}(I_{\sH_k}\otimes X_k)^\dag (I_{\sH_k}\otimes O_k)
(I_{\sH_k}\otimes X_k)(I_\sK\otimes C)]$ for any $C\in\Bnd{H}$, namely
the vanishing of
$\Tr\{\oplus_{k\in\set{S}}O_kX_k\Tr_{\sH_k}[P_k(I_\sK\otimes
C)P_k]X_k^\dag\}$ for any 
$C\in\Bnd{H}$, and upon defining $S=\oplus_{k\in\set{S}}O_k$,
the statement (\ref{Ocart2}) rewrites
\begin{equation}
\begin{split}
S\in&\oplus_{k\in\set{S}}\Bndd{\Rng(X_k)}, \\
&\Tr\{S\oplus_{k\in\set{S}}X_k\Tr_{\sH_k}[P_k(I_\sK\otimes\Bnd{H})P_k]X_k^\dag\}=0
\Longrightarrow S=0,
\end{split}
\end{equation}
namely, since the only operator in the linear space
$\oplus_{k\in\set{S}}\Bndd{\Rng(X_k)}$ orthogonal to the subspace
$\oplus_{k\in\set{S}}X_k\Tr_{\sH_k}[P_k(I_\sK\otimes\Bnd{H})P_k]X_k^\dag$
is the null operator, one has
\begin{equation}
\oplus_{k\in\set{S}}\Bndd{\Rng(X_k)}=\oplus_{k\in\set{S}}
X_k\Tr_{\sH_k}[P_k(I_\sK\otimes\Bnd{H})P_k]X_k^\dag.
\end{equation}
\qed
\begin{corollary}
As in Theorem \ref{th:extremalR}, a necessary condition for extremality is
\begin{equation}
\sum_{k\in\set{S}}\rnk(X_k)^2\le\dim(\sH)^2,\label{necessarymap1}\\
\end{equation}
\end{corollary}
\begin{corollary}
Any rank-one covariant QO is extremal.
\end{corollary}
\Proof For $\rnk(X)=1$ the set $\set{S}$ must contain only one equivalence
class, and the iff condition (\ref{iffmap2}) of Theorem \ref{th:extremalR} is then trivially satisfied.\qed  
\begin{corollary}
For an irreducible representation any extremal covariant QO must be rank-one.
\end{corollary}
\begin{corollary}[Choi]\label{c:choi}
In the non covariant case, a QO $\map{M}$ from
$\Bnd{K}$ to $\Bnd{H}$ is extremal iff it can be written in the form
$\map{M}(O)=\sum_i W_i^\dag O W_i $, with $W_i\in\Bnd{H,K}$ and the
set of operators $\{W_i^\dag W_j\}$ linearly independent.
\end{corollary}
\Proof The non covariant case corresponds to the trivial covariance
group $\gG=\gI$, \ie the group containing only the identity
element. This corresponds to have 
just a single equivalence class, with multiplicity equal to
$\dim(\sH\otimes\sK)$. Then, as in the proof of point {\bf 2.} of
Theorem \ref{th:extremalR} the extremality of
$R=X^\dag X\in\Bnd{H\otimes K}$ is equivalent to the injectivity
of the map $\map{W}(A)=\Tr_\sK[X^\dag AX]$ on
$\Bndd{\Rng(X)}$. According to Lemma \ref{l:partr2}, using the
singular value decomposition $X=\sum_i|V_i\>\< W_i|$, 
with $|V_i\>$ orthonormal basis for $\Rng(X)$ and
$|W_i\>\in\sK\otimes\sH$, one has
$\map{M}(O)=\sum_i W_i^\dag O W_i $ for $O\in\Bnd{K}$, and 
$\map{W}(A)=\sum_{ij}\< V_i|A|V_j\>\transp{W}_i W_j^*$ for
$A\in\Bndd{\Rng(X)}$, and injectivity of $\map{W}$ is equivalent to
linear independence of the set of operators $\{W_i^\dag W_j\}$.\qed 
\medskip
Corollary \ref{c:choi} is the same as Choi theorem \cite{Choi}.
Notice that differently from the case of QO's, for POVM's
the non covariant case cannot be recovered as a special case of the
covariant classification, since the group itself (or, more generally, the homogeneous factor
space) coincides with the probability space $\Klm$ of the POVM, whence
trivializing $\gG$ also trivializes $\Klm$.
\example\label{excl1} Consider the phase-covariant cloning\cite{opt_pcc,clon} for equatorial qubits
from 1 to 2 copies. This correspond to $\gG=\U{1}$, with
representations $U_\phi=e^{i\phi |1\>\< 1|_0}$ and $V_\phi=e^{i\phi
  \sum_{s=1}^2|1\>\< 1|_s}$ where $s=0$ denotes the input qubit and
$s=1,2$ the output ones. Here $\sH=\Cmplx^2$ and $\sK=\sH^{\otimes 2}$. We
first need to decompose the representation $V_\phi\otimes U_\phi^*$. This is
made of one-dimensional representations, with characters $e^{ik\phi}$,
with $k=-1,0,1,2$ and multiplicities $m_{-1}=1$,  $m_0=3$, $m_1=3$, and
$m_2=1$. The necessary condition (\ref{necessarymap1}) in the present
case becomes $\sum_{k\in\set{S}}\rnk(X_k)^2\le\dim(\sH)^2=4$, which
means that we can have either a single equivalence class with
$\rnk(X_k)\le 2$, or two equivalence classes with $\rnk(X_k)=1$
each. Orthonormal bases for the supporting spaces
$\sH_k\otimes\Cmplx^{m_k}\equiv\Cmplx^{m_k}$ of the $k$th 
equivalence class of irreducible representations are reported in Table
\ref{suppspaces1} as subset of an orthonormal basis for the tensor
product $\sK\otimes\sH$. 
\begin{table}[hb]
\begin{center}
\begin{tabular}{|c||c||}
\hline
\hline
$k$ & $|k_i\>\otimes|h_j\>$ \\ 
\hline
\hline
-1&$|001\>$\\
\hline
0&$|101\>,\,|011\>,\,|000\>$\\
\hline
1&$|100\>,\,|010\>,\,|111\>$\\
\hline
2&$|110\>$\\
\hline
\hline
\end{tabular}\end{center}
\caption{Orthonormal bases for the supporting spaces
$\sH_k\otimes\Cmplx^{m_k}\equiv\Cmplx^{m_k}$ of the $k$th 
equivalence class of irreducible representations for 1 to 2
phase-covariant cloning. The orthonormal basis are chosen as subsets
of an orthonormal basis for the tensor product $\sK\otimes\sH$.\label{suppspaces1}} 
\end{table}
\begin{table}[ht]
\begin{center}
\begin{tabular}{|c||c|c|c||}
\hline
\hline
$\set{S}\doteq\{k\}$ & $\left\{|\psi_l^{(k)}\>\right\}$ &
$\left\{|\psi_l^{(k')}\>\right\}$ &\\ \hline\hline
$\{-1,2\}$ & $|001\>$ & $|110\>$ &\\
\hline
$\{0,1\}$ & $a|000\>+b|011\>+c|101\>$
& $a'|111\>+b'|100\>+c'|010\>$ & $\begin{matrix}
|a|^2+|b'|^2+|c'|^2=1\\
|a'|^2+|b|^2+|c|^2=1\end{matrix}$\\
\hline
$\{0,-1\}$ & $|000\>+a|011\>+b|101\>$
& $c|001\>$ &$|a|^2+|b|^2+|c|^2=1$\\
\hline
$\{1,-1\}$ & $a|100\>+b|010\>+c|111\>$
& $d|001\>$ & $\begin{matrix}
|a|^2+|b|^2=1\\
|c|^2+|d|^2=1\end{matrix}$\\
\hline
$\{1,2\}$ & $a|100\>+b|010\>+|111\>$
& $d|110\>$ & $|a|^2+|b|^2+|d|^2=1$\\
\hline
$\{0,2\}$ & $a|000\>+b|011\>+c|101\>$
& $d|110\>$ & $\begin{matrix}
|a|^2+|d|^2=1\\
|b|^2+|c|^2=1\end{matrix}$\\
\hline
$\{0\}$ & $\tfrac{1}{\sqrt2}|101\>+\tfrac{1}{\sqrt2}|011\>,|000\>$ & &\\
\hline
$\{1\}$ & $\tfrac{1}{\sqrt2}|010\>+\tfrac{1}{\sqrt2}|100\>,|111\>$ & &\\
\hline
\hline
\end{tabular}\end{center}
\caption{Cloning from 1 to 2 copies: classification of operators $R=\sum_{k\in\set{S}}R_k=\sum_l
  |\psi_l^{(k)}\>\<\psi_l^{(k)}|$ satisfying the necessary condition.  
\label{classestab1}} 
\end{table}
\par The operators $R=\sum_{k\in\set{S}}R_k=\sum_l
  |\psi_l^{(k)}\>\<\psi_l^{(k)}|$ satisfying the necessary conditions
and the trace-preserving condition are reported in Table
\ref{classestab1}. It is easy to check that the case of $\rnk(X_k)=2$
which would be possible only for $k=0$ or $k=1$ doesn't satisfy the
iff condition (\ref{iffmap}). Therefore it is possible to have only
rank-one operators $X_k$. 
\par As a specific optimization problem, let's consider the maximization of the fidelity averaged
over the two outputs
\begin{equation}
\begin{split}
F&=\<\psi|\tfrac{1}{2}\{\Tr_1[\dual{\map{M}}(|\psi\>\<\psi|)]+\Tr_2[\dual{\map{M}}
(|\psi\>\<\psi|)]\}|\psi\>\\&=\Tr[\tfrac{1}{2}(I\otimes |\psi\>\<\psi|
+|\psi\>\<\psi|\otimes I)\dual{\map{M}}(|\psi\>\<\psi|)]
\end{split}
\end{equation}
and for equatorial qubits we can choose $|\psi\>=|+\>$, where $|\pm\>\doteq\frac{1}{\sqrt2}(|0\>\pm|1\>)$. Then the fidelity rewrites as
\begin{align}
F&=\Tr[WR_\map{M}],\\
W&=|+\>\<+|^{\otimes 3}+\tfrac{1}{2}(|-\>\<-|\otimes|+\>\<+|+|+\>\<+|\otimes|-\>\<-|)
\otimes|+\>\<+|.
\end{align}
One can see that $W$ is invariant for permutations over the output
copies, and, by construction, also all vectors in Table \ref{classestab1}
have the same symmetry. Due to the special form of the fidelity, the optimal
map (satisfying $\map{M}(I_\sK)=I_\sH$) is obtained for
$\set{S}=\{0,1\}$ with corresponding  rank-two operator $R_\map{M}$
given by
\begin{equation}
\begin{split}
R_\map{M}=&|\psi^{(0)}\>\<\psi^{(0)}|+|\psi^{(1)}\>\<\psi^{(1)}|,\\
|\psi^{(0)}\>-=&\tfrac{1}{\sqrt2}(|000\>+\tfrac{1}{\sqrt2}|011\>+\tfrac{1}{\sqrt2}|101\>),\\
|\psi^{(1)}\>=&\tfrac{1}{\sqrt2}(|111\>|+\tfrac{1}{\sqrt2}|100\>+\tfrac{1}{\sqrt2}|010\>).
\end{split}
\end{equation}
\example\label{excl2} Consider the phase-covariant cloning\cite{opt_pcc,clon} for equatorial qubits
from 1 to 3 copies. This correspond to $\gG=\U{1}$, with
representations $U_\phi=e^{i\phi |1\>\< 1|_0}$ and $V_\phi=e^{i\phi
  \sum_{s=1}^3|1\>\< 1|_k}$ where $s=0$ denotes the input qubit and
$s=1,2,3$ the output ones. Here $\sH=\Cmplx^2$ and $\sK=\sH^{\otimes 3}$. We
\begin{table}[hb]
\begin{center}
\begin{tabular}{|c||c||}
\hline
\hline
$k$ & $|k_i\>\otimes|h_j\>$ \\ 
\hline
\hline
-1&$|0001\>$\\
\hline
0&$|1001\>,\,|0101\>,\,|0011\>,\,|0000\>$\\
\hline
1&$|1000\>,\,|0100\>,\,|0010\>,\,|1101\>,\,|1011\>,\,|0111\>$\\
\hline
2&$|1100\>,\,|1010\>,\,|0110\>,\,|1111\>$\\
\hline
3&$|1110\>$\\
\hline
\hline
\end{tabular}\end{center}
\caption{Orthonormal bases for the supporting spaces
$\sH_k\otimes\Cmplx^{m_k}\equiv\Cmplx^{m_k}$ of the $k$th 
equivalence class of irreducible representations for 1 to 3
phase-covariant cloning. The orthonormal basis are chosen as subsets
of an orthonormal basis for the tensor product $\sK\otimes\sH$.
\label{suppspaces2}} 
\end{table}
first need to decompose the representation $V_\phi\otimes U_\phi^*$. This is
made of one-dimensional representations, with characters $e^{ik\phi}$,
with $k=-1,0,1,2,3$ and multiplicities $m_{-1}=1$,  $m_0=4$, $m_1=6$,
$m_2=4$, and $m_3=1$.  Orthonormal bases for the supporting spaces
$\sH_k\otimes\Cmplx^{m_k}\equiv\Cmplx^{m_k}$ of the $k$th 
equivalence class of irreducible representations are reported in Table
\ref{suppspaces2} as subset of an orthonormal basis for the tensor
product $\sK\otimes\sH$. Again, since $\dim(\sH)=2$, the necessary
condition (\ref{necessarymap1}) says that we can have only one equivalence
class $k$ with $\rnk(X_k)\le 2$, or two equivalence classes both with
$\rnk(X_k)=1$. In Ref. \cite{clon} it is shown that the map which
optimizes the averaged equatorial fidelity is actually given by the rank-one
map for $\set{S}=\{1\}$ with corresponding operator $R_\map{M}$ given by
\begin{equation}
\begin{split}
R_\map{M}=&|\psi^{(1)}\>\<\psi^{(1)}|,\\
|\psi^{(1)}\>=&\tfrac{1}{\sqrt3}(|1000\>+|0100\>+|0010\>+|1101\>+|1011\>+|0111\>).
\end{split}
\end{equation}
\bigskip
\par Notice that, as a consequence of the specific symmetric form of the chosen fidelity criterion, 
the cloning maps of the examples \ref{excl1} and \ref{excl2} are both symmetrical, namely the output
Hilbert space 
is indeed restricted to the symmetric tensor space $\left(\sH^{\otimes n}\right)_+$. Clearly,with
the same method also nonsymmetric types of cloning can be analyzed well.
\example Consider a generic covariant QO with $\sK\simeq\sH$,
$V_g=U_g$ and $\gG=SU(d)$, where $d=\dim(\sH)$. In this case the
representation $U_g\otimes U_g^*$ has two irreducible components: one
which is one-dimensional, corresponding to the invariant vector
$|I\>\in\sH^{\otimes 2}$, and one on the orthogonal complement, 
and the two components will be denoted by $k=0$ and $k=1$, respectively.
Since both the irreducible components of the representation have unit
multiplicity, the operator $R=X^\dag X$ must have
$X=\sum_{k\in\set{S}}c_kP_k$, $c_k\in\Cmplx$, $P_k$ denoting the
orthogonal projector on the invariant space of the irreducible component
$k$, and the necessary condition (\ref{necessarymap1}) is trivially
satisfied. On the other hand, one can see that the iff condition (\ref{iffmap}) is
satisfied for the irreducible representations $\set{S}=\{0\}$ and $\set{S}=\{1\}$,
whereas for the reducible one $\set{S}=\{0,1\}$ the map 
$\map{T}(O)=\Tr_\sK[X^\dag O X]$ is never
injective on $\aA_\gG'\cap\Bndd{\Rng(X)}$ (one has
$\Tr_\sK[X^\dag O X]=\frac{1}{d}[|c_0|^2a_0+(d^2-1)|c_1|^2a_1]I_\sH$
for $O=a_0P_0+a_1P_1$, $a_0,a_1\in\Cmplx$). Therefore, the only
trace-preserving optimal maps are those corresponding to the operators
$R=|I\>\< I|$ and $R=\frac{d}{d^2-1}(I^{\otimes 2} -\frac{1}{d}|I\>\<
I|)$, corresponding to the trivial map $\map{M}=\map{I}$ and to the
so-called isotropic depolarizing channel
$\map{M}(O)=\frac{d}{d^2-1}\Tr[O]I_\sH-\frac{1}{d^2-1}\rho$. 
Finally, notice that in the present example the optimal covariant maps
are compatible only with (multiple of) the trace-preserving condition,
since both partial traces $\Tr_\sK[P_k]$ are proportional to the identity.
\example We consider now the same problem as in the previous example, but now with
$V_g=U_g^*$. In this case we need to consider the positive operators
$R$ which are invariant under $U_g^*\otimes U_g^*$. It will be easier
to consider the representation $U_g\otimes U_g$ and then take the
complex conjugate of $R$ at the end. Now we have again two
irreducible inequivalent components, say $k=\pm$ with invariant spaces
$(\sH^{\otimes 2})_{\pm}$, the symmetric and the antisymmetric spaces.
As in the previous example, the general form of $R=X^\dag X$ is
$X=\sum_{k\in\set{S}}c_kP_k$, $c_k\in\Cmplx$,  and
$P_\pm=\frac{1}{2}(I_\sH^{\otimes 2}\pm E)$, 
where $E$ is the swap operator on the tensor product. However, the map
$\map{T}(O)=\Tr_\sK[X^\dag O X]$ is iniective on
$\aA_\gG'\cap\Bndd{\Rng(X)}$ only for representations with a single irreducible
component. One can see that $\Tr_\sH[P_\pm]=\frac{1}{2}(d\pm 1)I_\sH$,
and only trace-preserving (or multiplying by a constant) QO's  are
compatible with the present covariance. In conclusion,  
the only extremal covariant operators are $R_{\pm}=(d\pm 1)^{-1}(I^{\otimes
  2}\pm E)$, corresponding to the channels 
$\map{M}_\pm(O)=(d\pm 1)^{-1}[\Tr(O)I_\sH\pm\transp{O}]$. The map
$\map{M}_+$ is the optimal transposition map of Ref.\cite{opttransp}.
\section*{Acknowledgments}
\par I'm grateful to Koenraad Audenaert for having pointed to my
attention the work of Li and Tam\cite{LiTam94} which inspired the
present work. I'm also very grateful to Giulio Chiribella for
having pointed out an error in a preliminary version, and for a
careful reading of the present manuscript. A
substantial part of this work has been worked out during the {\em
Benasque Session on Quantum Information and Communication 2003}. 
It has been sponsored by INFM through the project PRA-2002-CLON, and
by EEC and MIUR through the cosponsored ATESIT project IST-2000-29681
and Cofinanziamento 2003.  Partial support is also acknowledged from
MURI program administered by the Army Research Office under Grant
No. DAAD19-00-1-0177. 


\begin{thebibliography}{}
\bibitem{note1} It must be noticed that in the present paper we are focusing only on the statistical
description of the measurement process---not on the change of state due to the measurement, nor on
its realization scheme---and for this reason we identify the measurement covariance with the
covariance of the POVM. The the state change due to the measurement is described by the so-called 
"instrument" \cite{davieslewis}, and the measuring scheme is given in terms of a unitary interaction
with an ancilla and an orthogonal measurement on it. Clearly, the POVM covariance does not imply the
covariance of all of its instruments, and the covarance of an instrument does not imply the
covariance of all of the measurement realizations of the instrument.
\end{thebibliography}
\end{document}